\documentclass[modern]{aastex61}
\shorttitle{High-resolution Study of Flow and Magnetic Field Evolution}
\shortauthors{Wang et al.}
\usepackage[T1]{fontenc}
\usepackage{textcomp}
\usepackage{colordvi}
\usepackage[normalem]{ulem}

\newcommand{\sm}{$\sim$}
\newcommand{\kms}{km~s$^{-1}$}

\begin{document}

\title{EVOLUTION OF PHOTOSPHERIC FLOW AND MAGNETIC FIELDS ASSOCIATED WITH THE 2015 JUNE 22 M6.5 FLARE}

\author{Jiasheng Wang}
\affiliation{Space Weather Research Laboratory, New Jersey Institute of Technology, University Heights, Newark, NJ 07102-1982, USA; jw438@njit.edu, chang.liu@njit.edu, na.deng@njit.edu, haimin.wang@njit.edu}
\affiliation{Big Bear Solar Observatory, New Jersey Institute of Technology, 40386 North Shore Lane, Big Bear City, CA 92314-9672, USA}
\affiliation{Center for Solar-Terrestrial Research, New Jersey Institute of Technology, University Heights, Newark, NJ 07102-1982, USA}

\author{Chang Liu}
\affiliation{Space Weather Research Laboratory, New Jersey Institute of Technology, University Heights, Newark, NJ 07102-1982, USA; jw438@njit.edu, chang.liu@njit.edu, na.deng@njit.edu, haimin.wang@njit.edu}
\affiliation{Big Bear Solar Observatory, New Jersey Institute of Technology, 40386 North Shore Lane, Big Bear City, CA 92314-9672, USA}
\affiliation{Center for Solar-Terrestrial Research, New Jersey Institute of Technology, University Heights, Newark, NJ 07102-1982, USA}

\author{Na Deng}
\affiliation{Space Weather Research Laboratory, New Jersey Institute of Technology, University Heights, Newark, NJ 07102-1982, USA; jw438@njit.edu, chang.liu@njit.edu, na.deng@njit.edu, haimin.wang@njit.edu}
\affiliation{Big Bear Solar Observatory, New Jersey Institute of Technology, 40386 North Shore Lane, Big Bear City, CA 92314-9672, USA}
\affiliation{Center for Solar-Terrestrial Research, New Jersey Institute of Technology, University Heights, Newark, NJ 07102-1982, USA}

\author{Haimin Wang}
\affiliation{Space Weather Research Laboratory, New Jersey Institute of Technology, University Heights, Newark, NJ 07102-1982, USA; jw438@njit.edu, chang.liu@njit.edu, na.deng@njit.edu, haimin.wang@njit.edu}
\affiliation{Big Bear Solar Observatory, New Jersey Institute of Technology, 40386 North Shore Lane, Big Bear City, CA 92314-9672, USA}
\affiliation{Center for Solar-Terrestrial Research, New Jersey Institute of Technology, University Heights, Newark, NJ 07102-1982, USA}

\begin{abstract}
The evolution of photospheric flow and magnetic fields before and after flares can provide important information regarding the flare triggering and back reaction processes. However, such studies on the flow field are rare due to the paucity of high-resolution observations covering the entire flaring period. Here we study the structural evolution of penumbra and shear flows associated with the 2015 June 22 M6.5 flare in NOAA AR 12371, using high-resolution imaging observation in the TiO band taken by the 1.6~m Goode Solar Telescope at Big Bear Solar Observatory, with the aid of the differential affine velocity estimator method for flow tracking. The accompanied photospheric vector magnetic field changes are also analyzed using data from the Helioseismic and Magnetic Imager. As a result, we found, for a penumbral segment in the negative field adjacent to the magnetic polarity inversion line (PIL), an enhancement of penumbral flows (up to an unusually high value of \sm2~\kms) and extension of penumbral fibrils after the first peak of the flare hard X-ray (HXR) emission. We also found an area at the PIL, which is co-spatial with a precursor brightening kernel, exhibits a gradual increase of shear flow velocity (up to \sm0.9~\kms) after the flare. The enhancing penumbral and shear flow regions are also accompanied by an increase of horizontal field and decrease of magnetic inclination angle(measured from the solar surface). These results are discussed in the context of the theory of back reaction of coronal restructuring on the photosphere as a result of flare energy release.
 
\end{abstract}


\section{Introduction} \label{sec:intro}

It is generally believed that solar flares are powered by free magnetic energy stored in the corona \citep{priest02}, and that the energy released via magnetic reconnection causes plasma heating and particle acceleration \citep{shibata11}, producing various flaring signatures at multiple heights in the solar atmosphere. Although tremendous efforts have been devoted to flare studies, many fundamental physical problems are still not well understood, such as the energy build-up and triggering of flares, and the consequent back reaction into the low atmosphere as a result of the sudden coronal restructuring \citep[see][for a review]{wang15}. Since coronal magnetic fields are anchored to the dense photosphere, insights into these problems can be obtained by studying the structural evolution of photospheric magnetic field and the closely coupled plasma flow field leading to and from flare events. In general, the first problem above is often related to relatively long-term (in hours to days) evolution of magnetic and flow fields (e.g., emerging fluxes, shear and converging flows), while the second problem above deals with short-term (in tens of minutes) flare-induced changes of magnetic/flow fields down to the photosphere of solar eruption.

It is particularly noticeable that back reaction is traditionally considered to be small because of the large inertia of the photosphere. The reconfiguration of coronal field is the focus of almost all models of flares and the often associated coronal mass ejections (CMEs), which generally do not consider the restructuring of magnetic and flow fields in the photosphere due to the assumed line-tying effect. Based on the principles of energy and momentum conservation, \citet{hudson08} and \citet{fisher12} quantitatively investigated the back reaction on the solar surface and interior as a result of the coronal field evolution (specifically, implosion) after energy release, and pointed out that flares/CMEs would make the photospheric magnetic field become more horizontal at the flare-related magnetic polarity inversion lines (PILs). The authors formulated the resulted changes of the integrated vertical (downward) and horizontal Lorentz force exerted on the photosphere from the corona, and suggested that the former may energize seismic waves and the latter may cause plasma bulk motions. Thus far, these predictions of the back reaction theory have been substantiated by many aspects of observational results of flare-related photospheric magnetic/flow field changes, including the rapid and permanent step-like increase of horizontal field at flaring PILs \citep[e.g.,][]{wang92,liu05,wangj09,wang10,liu12,sun12,petrie12,sun2017}, possible linkage to the excitation of seismic waves \citep[e.g.,][]{alvarado12,liu14}, and sunspot displacement and rotations \citep[e.g.,][]{anwar93,wangs14,liu16}. The downward collapse of coronal current systems presumably due to coronal implosion following flares/CMEs is also corroborated by time sequence of nonlinear force-free field models \citep[e.g.,][]{sun12,liu14} and is reflected in MHD simulations \citep[e.g.,][]{fan05,fan10}.

Besides magnetic field variations, changes of sunspot structure in white light have also been found to be consistent with the back reaction scenario. Notably, sunspot penumbrae are a direct indication of horizontal photospheric field, with their fibrils following the direction of magnetic azimuth \citep{solanki03}. The outward plasma flows along fibrils (known as Evershed flows) can have a velocity up to \sm4~\kms\ in photosphere from spectroscopic observations \citep[e.g.,][]{solanki94}, and can reach up to \sm8~\kms\ in chromosphere \citep{balthasar14}, while optical penumbral flows measured by flow tracking methods based on imaging observations generally result in a velocity on the order of \sm1~\kms\ \citep[e.g.,][]{tan09, verma12}. The penumbral structure and its carried flows are governed by the magnetic field strength and especially the field inclination, as revealed by previous observations \citep[e.g.,][]{ichimoto10,deng11c} and MHD simulations \citep{rempel09,kitiashvili09}. Thus, the expected more horizontal photospheric field configuration in response to the coronal restructuring as described above is strongly evidenced by observations of flare-related darkening and/or formation of penumbral structure, together with strengthened horizontal field and decreased inclination angle, near central flaring PILs \citep{liu05,deng05,chen07b,li09,wang13,xuz16}. It can be noted that these papers studied the structural evolution of penumbra mainly based on the overall intensity of penumbral segments, without examining in detail the associated penumbral flows. Understandably, studying the flare-related flow field evolution requires high-quality, high-resolution observations that cover the entire flare interval, which are, however, relatively rarely available. Using high-resolution (0$\farcs$2) G-band images from Hinode, \citet{tan09} observed enhancement of the central penumbral region of the $\delta$ spot in NOAA AR 10930 associated with the December 13 X3.4 flare. By employing the local correlation tracking (LCT) technique, the authors were able to detect shear flows (i.e., opposite-directed flows at the two sides of the PIL) in the penumbral region, which exhibited a significant decrease associated with the flare, probably due to the magnetic restructuring and energy release. Importantly, shear flows can contribute to the build up of magnetic nonpotentiality in flaring regions \citep{harvey76,amari00,welsch09} and thus flare triggering, as in high resolution they show a close spatial proximity to the initial flare kernels \citep{yang04,deng06}. It is obvious that more studies of both the flow and magnetic field evolution in penumbrae can help to shed further light on our understanding of the photosphere-corona coupling related to flaring activities.

In this paper, we take advantage of the unprecedented resolution of the recently commissioned 1.6~m Goode Solar Telescope \citep[NST;][]{goode10a,cao10,goode12,varsik14} at Big Bear Solar Observatory (BBSO) to investigate the photospheric flow field associated with the 2015 June 22 M6.5 flare, and also study the corresponding magnetic field using data from the Helioseismic and Magnetic Imager \citep[HMI;][]{schou12} on board the Solar Dynamics Observatory (SDO). We concentrate on the structural evolution of penumbra near the flaring PIL, and make quantitative characterizations of the flare-related changes of the flow and magnetic properties. We compare our results with previous studies and discuss their implications in the context of the back reaction theory. The structure of this paper is as follows. We introduce the observations and data analysis methods in Section~\ref{data}, and describe the observational results in Section~\ref{results}. Major findings are summarized and discussed in Section~\ref{summary}.

\section{OBSERVATIONS AND DATA PROCESSING}\label{data}
With excellent seeing condition, BBSO/GST achieved diffraction-limited imaging on 2015 June 22 during \sm16:25--22:50~UT, thanks to the high-order AO-308 system (with 308 sub-apertures) and speckle image reconstruction. The obtained multiwavelength observations have revealed many interesting properties of the fully covered M6.5 flare, including flare precursors, sunspot rotation, and various fine structures \citep{wang17,liu16,jing16}. The essential data used in this study for tracking the photospheric flows are the images taken by the GST's Broad-Band Filter Imager at the TiO band (7057~\AA, 10~\AA\ bandpass), a proxy for the continuum photosphere, using a 2048~$\times$~2048 pixels CCD camera with a \sm70\arcsec\ field of view (FOV). The spatial resolution (at diffraction limit $\theta= \lambda / D$) of TiO images is 0$\farcs$09, and the temporal cadence is 15~s. As for magnetic field, we used 135 second cadence vector magnetograms from HMI's full-disk vector field data product \citep{hoeksema14,sun2017}, which is provided by the Joint Science Operations Center. The pixel scale of HMI data is 0$\farcs$5. The TiO images and magnetograms are aligned based on sunspot and plage features, with an alignment accuracy within \sm0$\farcs$3, which is the best accuracy by using interpolation. Also used are the 25--50~keV hard X-ray (HXR) time profile from the Fermi Gamma-Ray Burst Monitor \citep[GBM;][]{meegan09} for studying the timing of flare energy release, and 1700~\AA\ continuum (5000~K) images from the Atmospheric Imaging Assembly \citep[AIA;][]{lemen12} on board SDO for disclosing the flare precursor brightenings in the low atmosphere.

To derive the flow field on photosphere based on the TiO observation, we aligned TiO images to sub-pixel precision, normalized the intensity to that of a quiet-Sun area, and applied a 2~$\times$~2 image binning to increase the S/N ratio; we then applied the differential affine velocity estimator \citep[DAVE;][]{schuck06}, which is a demonstrated technique for flow detection and tracking. Here we set the tracking window to 25 pixels, trying to include enough structure information and at the mean time achieving a good resolution. We estimate quantitative uncertainty by analyzing data with tracking window range from 20 to 30 pixels, which results in a maximum relative standard deviation (RSD) as 5\%. For a validity check, we repeated the flow tracking using the LCT method \citep{november88}, which produced similar results. To minimize the effects of atmospheric seeing and five-minute period photospheric oscillation, a five minute running average was further made to the derived DAVE velocity vectors.

\begin{figure}[t]
\epsscale{1.2}
\plotone{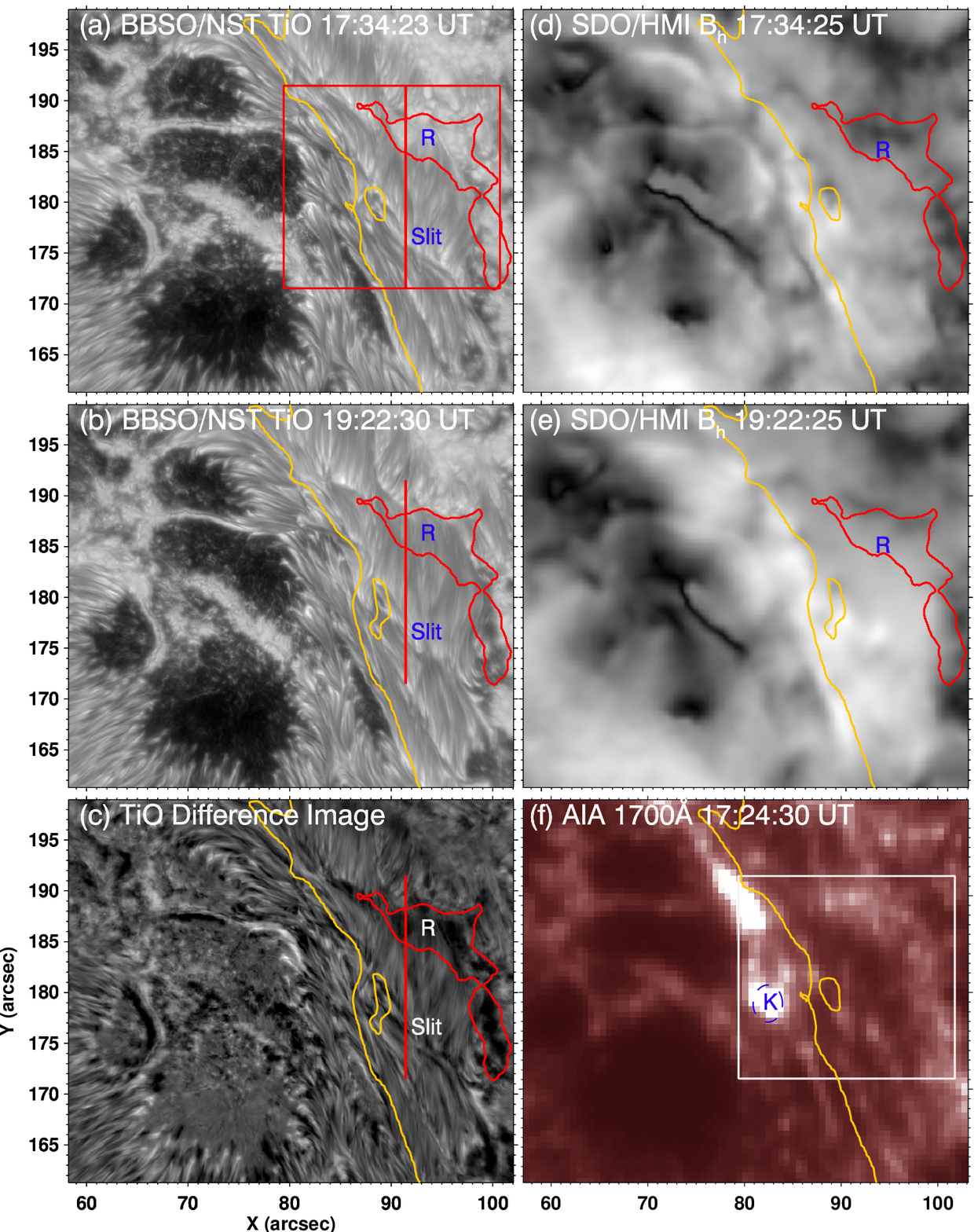}
\caption{Event overview. (a--c) Pre-flare, post-flare, and difference (19:22:30~UT frame subtract 17:34:23~UT frame) BBSO/GST TiO images, overlaid with the red contour indicating the newly formed penumbral region R determined based on the difference image. The vertical red line marks the slit used for the space-time slice image shown in Figure~\ref{f3}. (d--e) Pre- and post-flare SDO/HMI horizontal magnetic field maps, also overplotted with the contour of region R. (f) Pre-flare SDO/AIA 1700~\AA\ image, showing flare precursor K. The yellow contours superimposed in all panels represent the co-temporal PILs. The boxed region in (a) and (f) indicates the FOV of Figure~\ref{f2}. An animation of this figure is available in the online journal. \label{f1} }
\end{figure}


\begin{figure}[t]
\epsscale{1.2}
\plotone{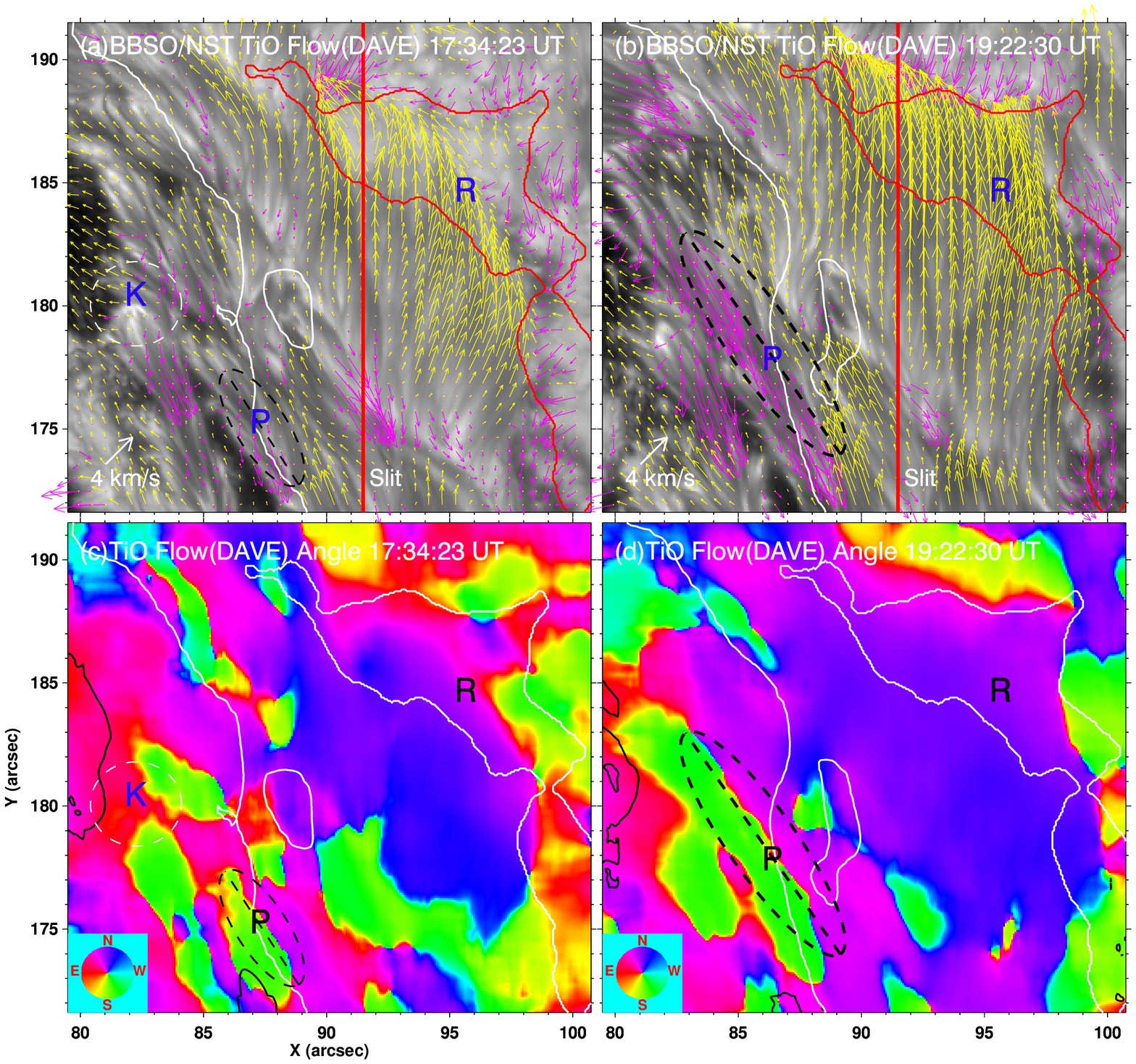}
\caption{Flow field in BBSO/GST TiO band. (a) and (b) Pre-flare (at 17:34:23~UT) and post-flare (at 19:22:30~UT) TiO images overplotted with arrows illustrating the flow vectors derived with DAVE. For clarity, arrows pointing northward (southward) are coded yellow (magenta). The white and red contours denote the PIL and the region R (see Figure~\ref{f1}), respectively. (c) and (d) Azimuth maps of corresponding flow vectors in (a) and (b), also overplotted with the PIL, precursor kernel, and region R contours. The shear flow region P showing the most obvious flare-related enhancement is outlined using the dashed ellipse, with its major axis quasi-parallel to the PIL. An animation of this figure is available in the online journal. \label{f2} }
\end{figure}


\section{ANALYSIS AND RESULTS}\label{results}
The flare-productive NOAA AR 12731 appears in the $\beta\gamma\delta$ configuration on 2015 June 22. The M6.5 flare of interest occurred in this region starts at 17:39~UT in GOES 1--8~\AA\ soft X-ray flux, and shows three main peaks in Fermi 25--50~keV HXR flux at 17:52:31, 17:58:37, and 18:12:25~UT \citep{liu16}. The separating two ribbons of the flare originate from penumbral regions very close to the PIL, and subsequently sweep across the main sunspot umbrae of opposite magnetic polarity \citep{wang17,jing16}. In this work we focus on the evolution of the central penumbral region of this $\delta$ configuration from the pre- to post-flare states. Dynamic details can be seen in the accompanying TiO animation in the online journal. Similar to previous studies, stepwise changes of physical properties are observed and are quantified by fitting to a step function \citep{sudol05}

\begin{equation} \label{eq1}
B(t)=a+bt+c\{1+\frac{2}{\pi}{\rm tan}^{-1}[n(t-t_0)]\} \ ,
\end{equation}

\noindent where $a$ and $b$ describe the strength and evolution of the background, $t$ is time, $c$ represents half amplitude of the step, $n$ controls the slope of the step, and $t_0$ is the middle point of the step. In this equation the time of the start of the change is $t_0-0.5\pi n^{-1}$.

\begin{figure}[t]
\epsscale{1.25}
\plotone{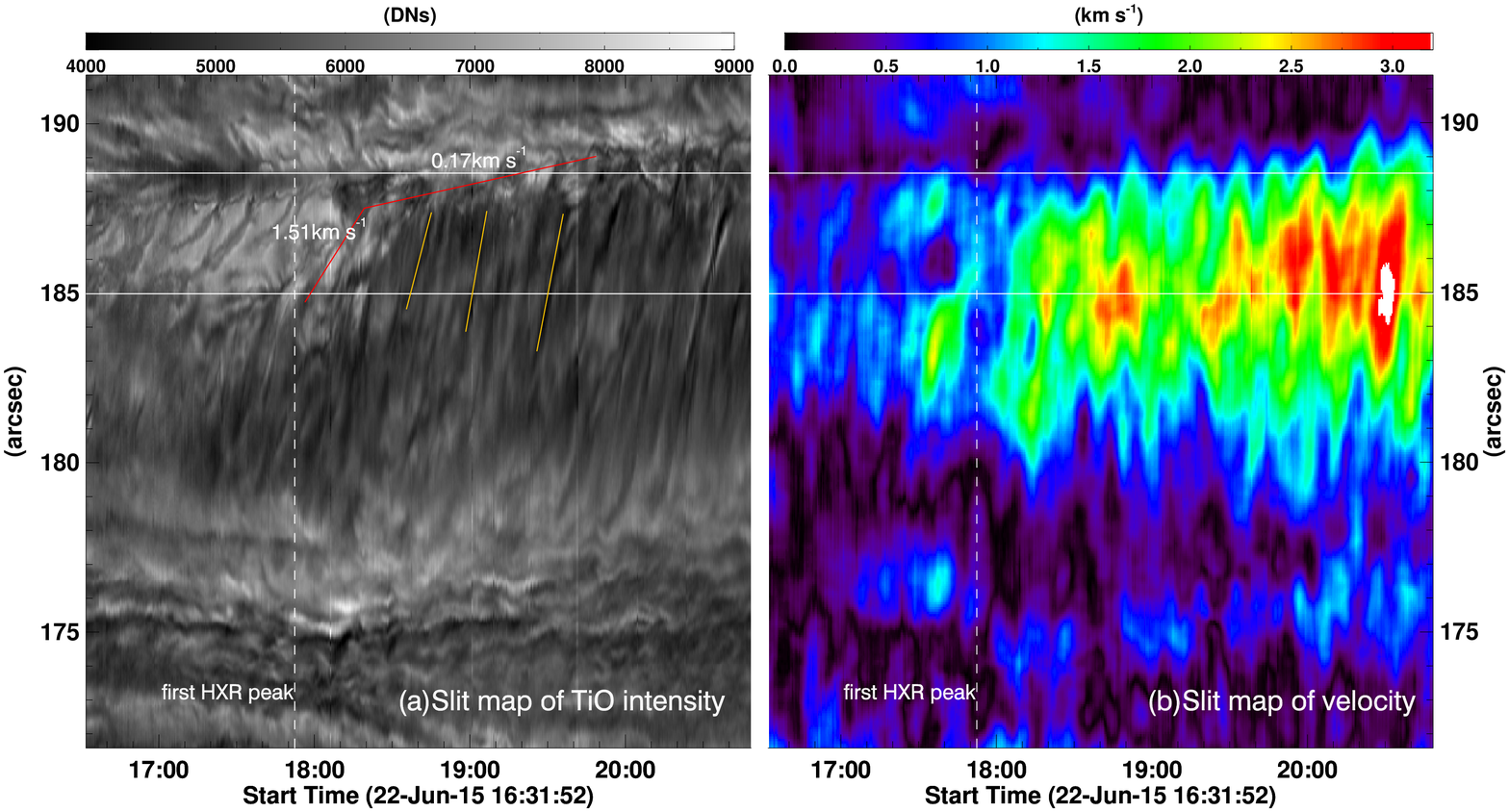}
\caption{Space-time slice image based on BBSO/GST TiO intensity images (a) and the derived time sequence of DAVE velocity field (b). The slit is marked as the red vertical line in Figures~\ref{f1}(a)--(c). The vertical dashed line denotes the time of the first HXR peak at 17:52:31~UT, the horizontal solid line denotes intersections of slit with region R in Figure~\ref{f2}(a) (b). The overplotted red lines in (a) trace several prominent evolving penumbral fibril features, and are used to estimate the flow velocity based on their slope. \label{f3} }
\end{figure}

\subsection{Evolution of Growing Penumbra Region}\label{penumbra}
By observing the time-lapse TiO movie, one can easily identify the most prominent flare-related penumbral evolution, that is, one segment of the central penumbra lying in the negative field becomes much enhanced (in terms of flow dynamics) and extends to the north, with a close temporal relationship with the flare. This can be readily seen by comparing the pre- (17:34~UT) and post-flare (19:22~UT) TiO images as shown in Figures~\ref{f1}(a) and (b), respectively. The difference image in Figure~\ref{f1}(c) (post-flare minus pre-flare state) displays a main darkened region R (encompassed by the red contour), which corresponds to the newly formed portion of the penumbra that was occupied by photospheric granulations. In Figure~\ref{f2}, we show the maps of flow vectors and azimuth derived with DAVE in the pre- and post-flare states. It is evident that after the flare, the flow vectors strengthen vastly, not only in the northern newly formed penumbral region R but also in the previously existing, common penumbral area; moreover, the azimuth of flows become more uniform, predominantly toward the north direction. An accompanying gradual increase of penumbral flow speed is shown in time-lapse flow movie. A similar overall enhancement of horizontal magnetic field across the existing and newly formed penumbra regions is also observed (cf. Figures~\ref{f1}(d) and (e)).

\begin{figure}[t]
\epsscale{1.2}
\plotone{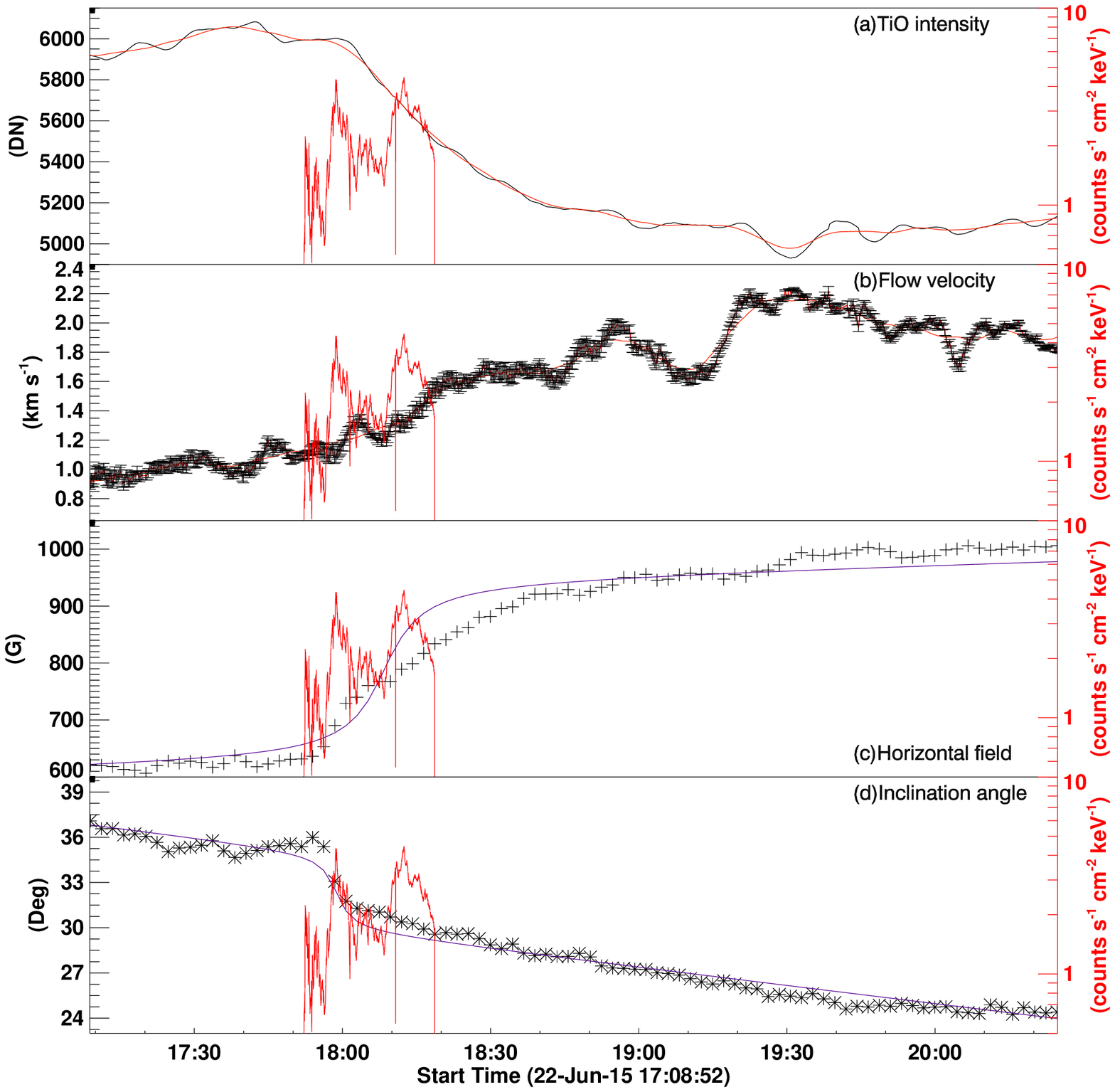}
\caption{Temporal evolution of physical properties (mean values) in the growing penumbral region R. (a) TiO intensity. (b) FLow velocity. (c) Horizontal magnetic field strength. (d) Magnetic inclination angle. In (c) and (d), data are fit to a step function (see Equation~\ref{eq1}). The overplotted magenta curve represents the Fermi HXR 25--50~keV flux. \label{f4} }
\end{figure}

To better depict the temporal evolution of the penumbral structure, in Figure~\ref{f3}(a) we construct the space-time slice image for a slit (marked as the red line in Figures~\ref{f1}(a)-(c)) along the penumbral growing direction, based on the TiO intensity images, and estimate speed of segmental motions of penumbra. We see that this penumbral segment begins to grow rapidly toward north from around the time of the first HXR peak (\sm17:52~UT), with its northern end of fibrils extending for a distance of 2$\farcs$6($\pm$$\farcs$3) in about 20 minutes (from which we found that the north end of fibrils extends at an average speed of 1.5($\pm$0.05)~\kms); later on, it continues to slowly extend northward for about two hours (at an average speed of 0.17($\pm$0.01)~\kms). We also trace several prominent evolving fibril features of the main penumbral body using red lines in Figure~\ref{f3}(a), which exhibit that after the first HXR peak, the penumbral flows become more conspicuous, with a velocity reaching up to 2.3($\pm$0.12)~\kms\ (estimated based on the slope of red lines). The evolution of flow velocity is also directly manifested in Figure~\ref{f3}(b), which is the space-time slice image for the same slit but built upon the time sequence of the velocity field derived with DAVE. The penumbral flows strengthens from an average velocity of 0.8($\pm$0.04)~\kms\ at 30 minutes before the first HXR peak to 2.2($\pm$0.11)~\kms\ at 90 minutes afterwards within the region R edges, which is comparable with the velocity estimation using the time slices. It is pertinent to point out that this velocity (2.2($\pm$0.11)~\kms) of penumbral flows is about twice as large as that measured based on Hinode G-band images \citep{tan09}, presumably due to (1)the higher resolution of the TiO data used in the present analysis, and (2) intrinsic property of this specific active region we studied.

In Figure~\ref{f4}, we further plot time profiles of mean TiO intensity, flow velocity, horizontal field strength, and magnetic inclination angle (defined as the angle of magnetic field vector with respect to solar surface) of the newly developed penumbral region R, and compare their timings with the HXR emission. The results show that after about the first HXR peak, the region R begins to show an increase of penumbral flow velocity together with a penumbral darkening, an increase of horizontal magnetic field strength, and a decrease of field inclination angle. Specifically, according to the step function fittings, starting from 18:00~UT the intensity decreases 15($\pm$2)\% in 30 minutes; meanwhile, the horizontal field increases 150($\pm$15)~G in \sm18~minutes from 17:50 UT to 18:08 UT, with inclination angle decreasing 5($\pm$0.5)$^{\circ}$ in the same time period. Then horizontal field continues with a gradual increase of 200($\pm$20)~G from 18:08 UT to 19:40 UT, while inclination angle decreases 6($\pm$0.6)$^{\circ}$ for that time. In contrast, the flow velocity evolves more gradually, leading to an increase of 1.2($\pm$0.1)~\kms\ from 30 minutes before the first HXR peak to 90 minutes afterwards.

\begin{figure}[t]
\epsscale{1.2}
\plotone{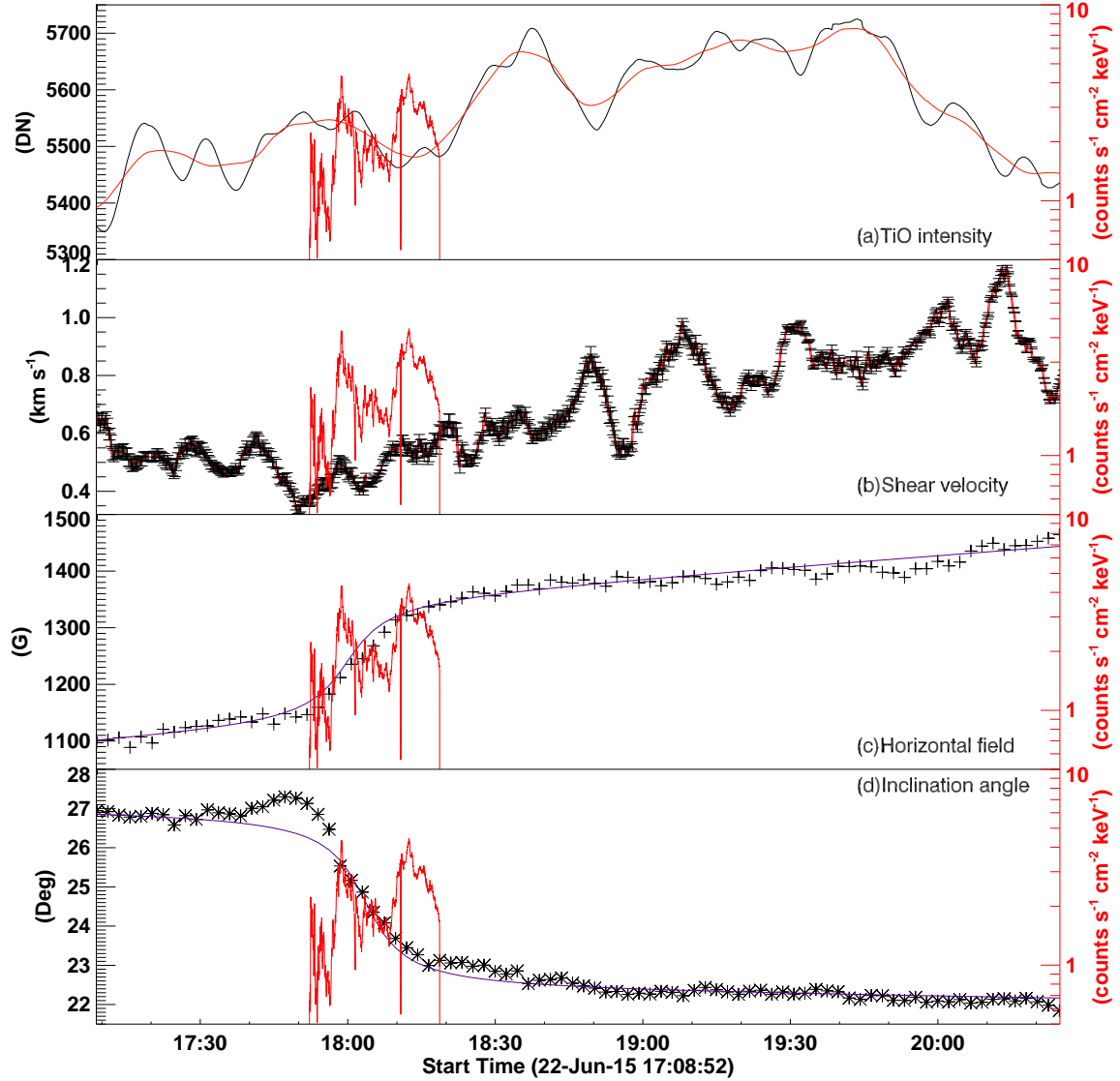}
\caption{Temporal evolution of physical properties (mean values) in the shear flow region P (the elliptical region in Figure~\ref{f2}d). (a) TiO intensity. (b) Shear flow velocity. (c) Horizontal magnetic field strength. (d) Magnetic inclination angle. In (c) and (d), data are fit to a step function (see Equation~\ref{eq1}). The overplotted magenta curve represents the Fermi HXR 25--50~keV flux. \label{f5}}
\end{figure}

\subsection{Evolution of Shear Flow near PIL}\label{sub:sff}
By examining the time sequence of TiO flow maps, we also find pronounced shear flows along the PIL in this flaring region. As can be seen in Figure~\ref{f2}(a), around region P the flows in the eastern side of the PIL pointing toward southwest while those in the western side of the PIL pointing toward northeast, constituting a clear shear flow pattern. In the azimuth map computed using the derived flow vectors (Figure~\ref{f2}(c)), the shear flow pattern can be recognized as a green-magenta feature with the central dividing line running along the PIL. We note that (1) this shear flow region is co-spatial with a ``magnetic channel'' structure (with multiple polarity inversions) that can be identified using high-resolution magnetic field observations for this region \citep{wang17}, and is also adjacent to a low-atmospheric, flare precursor brightening kernel K (cf. Figures~\ref{f1}(f) and \ref{f2}(a)(c)). Such a spatial correlation between shear flows and initial flare kernels was also found before using high-resolution observations \citep{yang04,deng06}, suggesting that the shear flow may contribute to the flare triggering process. (2) Most intriguingly, the shear flow enhances and its region expands substantially after the flare (cf. Figures~\ref{f2}(a)(c) and (b)(d)). For a quantitative analysis, we define the magnitude of shear flow velocity as $v_{\rm shear} = \overline{v}_{\rm pos} - \overline{v}_{\rm neg}$, where $\overline{v}_{\rm pos}$ ($\overline{v}_{\rm neg}$) is the flow velocity in the positive (negative) field region, in the direction parallel to the PIL. For the results in this paper, positive value of $v_{\rm shear}$ represents counterclock-wise direction of shear. In Figure~\ref{f5}(b), we plot the time profile of the shear flow velocity averaged over the region P (in Figure~\ref{f2}(d)). A gradual increase of shear flow is seen from the first HXR peak, and later reaches 0.9($\pm$0.05)~\kms\ at 19:30 UT. This magnitude of shear flows is 50($\pm$5)\% larger than that of the maximum shear flows in the event of \citet{tan09}.

The shear flow region P has an insignificant intensity variation (within 5\%) associated with the flare (Figure~\ref{f5}(a)). Nevertheless, it still shows appreciable, stepwise changes of magnetic properties. Fittings using a step function indicate that starting from 17:50~UT (co-temporal with the first HXR peak), the horizontal field strength increases 220($\pm$20)~G from 17:50 UT to 18:20 UT (Figure~\ref{f5}(c)), while the field inclination angle decreases 5($\pm$0.5)$^{\circ}$ from 17:50 UT to 18:20 UT (Figure~\ref{f5}(d)). Compared to the growing penumbral region R, the transition time of horizontal field  changes in shear region P is longer and followed by a flater gradual increase.

\section{SUMMARY AND DISCUSSIONS}\label{summary}
We have presented a detailed study of the structural evolution of photospheric flow and magnetic fields associated with the 2015 June 22 M6.5 flare, concentrating on the central penumbrae of the $\delta$ region around the flaring PIL. We tracked the penumbral flow field with the DAVE method using high-resolution TiO images from BBSO/GST, and also analyzed the associated magnetic field using SDO/HMI vector magnetograms. The main results are summarized as follows.

\begin{enumerate}

\item Beginning from the first HXR peak, one segment of penumbra lying in the negative field experiences a distinct grow, extending northward at 1.5($\pm$0.05)~\kms\ for a distance of 2$\farcs$6($\pm$$\farcs$3); meanwhile, the flow velocity within the entire penumbral region becomes more pronounced, gaining a 115($\pm$10) \% increase reaching up to 2.2($\pm$0.11)~\kms\ at \sm90~minutes after first flare peak. These structure and flow field evolutions are accompanied by a step-like increase of horizontal magnetic field by 150($\pm$15)~G and decrease of inclination angle by 5($\pm$0.5)$^{\circ}$ in 18 minutes from 17:50 UT to 18:08 UT.

\item A region of shear flow at the flaring PIL is found next to the location of a flare precursor brightening, and expands significantly after the flare. From the first HXR peak, the shear flow velocity increases gradually by 0.4($\pm$0.1)~\kms\ from 0.5($\pm$0.05)~\kms\ at 17:30 UT to 0.9($\pm$0.05)~\kms\ at 19:30 UT. As for magnetic field properties in the shear flow region, the horizontal field strength increases 220($\pm$20)~G while the inclination angle decreases 5($\pm$0.5)$^{\circ}$ from 17:50 UT to 18:20 UT, both following a stepwise fashion. In comparison with the extending penumbral segment, the transition time of step-like magnetic field changes in the shear flow region is \sm10~minutes longer. 

\end{enumerate}

Our results of the strengthening of penumbral structure at the center of $\delta$ region, the associated increase (decrease) of horizontal field (inclination angle), and their close timing relationship to the flare energy release strongly favor the back reaction of coronal fields as the underlying cause, which results in a more horizontal configuration of photospheric field. While most other work only deals with the flare-related overall intensity change of penumbral structure, we are able to study the evolution of penumbral flow field owing to the high-resolution BBSO/GST data. We show that although noticeable intensity darkening is only observed in the newly formed penumbral portion, the strengthening of penumbra is predominantly manifested as the enhancement of penumbral flow velocity and enlargement of penumbral area. The inconsistency of our results with the claim of darkening of central penumbral feature in some previous studies may due to the limitation of the used low-resolution images \citep[e.g.,][]{liu05} and also the complexity of the central penumbral area near PIL \citep{wang17}. 

The also revealed spatial correlation between the shear flow and flare precursor brightening kernel corroborates the importance of photospheric flow field in triggering flares. In this event, the precursor brightenings are caused by the reconnection between emerging fluxes in the magnetic channel with ambient large-scale sheared loops \citep{wang17}. The produced lower-lying fields near the PIL could be readily subject to the downward collapse of coronal fields \citep{liu12}, which may explain the observed more rapid magnetic field changes in the shear flow region. We also remark that the change of shear flows associated with flares is an observational fact of particular interest but its study has been lacking. Like penumbral enhancement, we also tend to attribute the observed increase of shear flow after the present flare to the more horizontal photospheric fields due to back reaction. An increase of shear flows near the PIL was also found by \citet{deng06} associated with an X10 flare in NOAA AR 10486, although the shear flows are not located in the central penumbra with accompanied magnetic field changes. To our knowledge, the study most closely related to our work was conducted by \citet{tan09} on an X3.4 flare in NOAA AR 10930. Contrarily, the authors saw a rapid decrease of shear flow in the central penumbral region, and explained it as a signature of magnetic energy relaxation between the two major magnetic polarities of the $\delta$ spot, one of which shows a continuous rotation before the flare. Such a relaxation could be viewed in line with the finding of sudden release of magnetic shear in the horizontal direction along the PIL of $\delta$ spots, as reported by \citet{wang06}. We speculate that the evolution of shear flows, due to their special locations along the flaring PILs, could be affected by both the back reaction from the above corona and the response of photospheric magnetic polarities to the energy release. The back reaction creates sheared flux system near the surface \citep{wang94} that may be reflected as increased shear flow, while relaxation of large scale system may reduce shear flow.

In summary, high-resolution observations of photospheric flow field evolution from pre- to post-flare stages, together with the analysis of magnetic field changes, will be valuable in advancing our understanding of flare triggering and back reaction processes. Further studies in this direction are highly desired and would certainly contribute to the science preparation in high-resolution flare studies for the upcoming 4~m Daniel K. Inouye Solar Telescope. 

\acknowledgments
We are grateful to the anonymous referee for valuable comments that help improving the paper significantly.

We thank the BBSO, SDO, and Fermi teams for providing the data. This work is supported by NASA under grants NNX13AF76G, NNX13AG13G, NNX16AF72G, and 80NSSCITK0016, and by NSF under grants AGS 1348513 and 1408703. The BBSO operation is supported by NJIT, NSF under grant AGS 1250818, and the GST operation is partly supported by the Korea Astronomy and Space Science Institute and Seoul National University and by the strategic priority research program of CAS with Grant No. XDB09000000. This work uses the DAVE code written and developed by P. W. Schuck at the Naval Research Laboratory.

\facilities{Big Bear Solar Observatory, SDO}

\bibliographystyle{aasjournal}
\bibliography{manuscript}                                                      

\end{document}